\documentclass[preprint,english,aps,prl]{revtex4}
\usepackage[T1]{fontenc}
\usepackage[latin9]{inputenc}
\usepackage{prettyref}
\usepackage{float}
\usepackage{amsmath}
\usepackage{graphicx}

\providecommand{\tabularnewline}{\\}

\usepackage{babel}

\begin{document}

\title{Dislocation Mobility in a Quantum Crystal: the Case of Solid $^{4}\mathrm{He}$}

\author{Renato Pessoa}

\email{rpessoa@ifi.unicamp.br}

\author{S. A. Vitiello}

\email{vitiello@unicamp.br}

\author{Maurice de Koning}

\email{dekoning@ifi.unicamp.br}

\affiliation{Instituto de F\'isica Gleb Wataghin, Caixa Postal 6165, Universidade
Estadual de Campinas - UNICAMP 13083-970, Campinas, SP, Brazil }

\date{\today}
\begin{abstract}
We investigate the structure and mobility of dislocations in \emph{hcp}
$^{4}\mathrm{He}$ crystals. In addition to fully characterizing the
five elastic constants of this system, we obtain direct insight into
dislocation core structures on the basal plane, which demonstrates
a tendency toward dissociation into partial dislocations. Moreover,
our results suggest that intrinsic lattice resistance is an essential
factor in the mobility of these dislocations. This insight sheds new
light on the possible correlation between dislocation mobility and
the observed macroscopic behavior of crystalline $^{4}\mathrm{He}$.
\end{abstract}


\maketitle
Dislocations are line defects that play a central role in the mechanical
deformation behavior of crystalline solids \cite{Hirth1982}. Their
activity is widely known to be pivotal in classical solids, controlling
phenomena such as fracture and brittle-ductile transitions in metals
and semiconductors. Much less is known \cite{Balibar2009}, however,
about the influence of dislocations on the properties of quantum crystals,
which are solids where the quantum-mechanical zero-point kinetic energy
is significant compared to the typical energy scale of the interatomic
interactions  \cite{Anderson1984}.

Recent experimental studies on crystalline solid $^{\text{4}}\mathrm{He}$,
the propotypical quantum solid, indirectly indicate that dislocations
are indeed involved in macroscopic phenomenona. The occurrence of
apparent superfluidity \cite{Kim2004a,Kim2004} and the observation
of elastic stiffening \cite{Day2007,Day2009,West2009a}, for instance,
have been linked to the mobility of dislocations. However, in contrast
to classical solids, for which an abundant body of experimental results
exists, the lack of specific experimental data concerning the behavior
of dislocations prevents a direct investigation of this relationship.
In view of these difficulties, one needs to resort to theoretical
approaches. In principle, a realistic picture of dislocations and
their properties is possible using \textsl{\emph{path-integral}}\textsl{
}or variational Monte Carlo simulations. However, given that dislocation
modeling requires a simultaneous treatment of different length scales
\cite{Bulatov2006}, one associated with the core properties and the
other associated with distances large compared to the atomic scale,
the viability of these direct approaches remains a subject of debate
\cite{Anderson2009,Boninsegni2007,Pollet2007}. 

In this Letter, adopting a methodology that incorporates both of the
relevant scales, we obtain fundamental insight into the properties
of dislocations in the archetypal quantum crystal: solid \emph{hcp}
$^{\text{4}}\mathrm{He}$. We compute the intrinsic dislocation structure
and mobility of four different dislocation types using the multiscale
paradigm of the semi-discrete Peierls-Nabarro (PN) model \cite{Bulatov1997,Lu2000,Lu2003}.
It constitutes a hybrid continuum-atomistic approach that captures
the long-range elastic fields as well as the lattice-discreteness
effects associated with the dislocation core. All parameters in the
model are determined using quantum-mechanical expectation values for
$^{\text{4}}\mathrm{He}$ applying the shadow wave function (SWF)
formalism \cite{Vitiello1988,MacFarland1994}. In addition to providing
key information concerning the elastic properties, our results shed
new light onto the possible role of dislocations in the experimental
observation of the elastic stiffening of solid and its potential connection
to the apparent superfluidity in this quantum crystal. 

\begin{figure}[H]
\includegraphics[width=9cm]{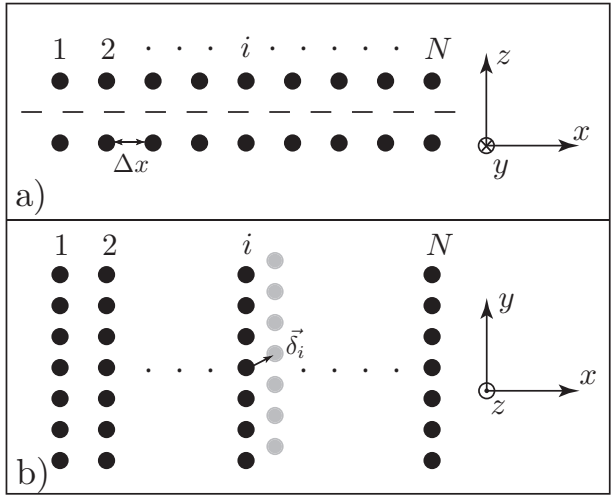}

\caption{\label{fig:Fig1}Degrees of freedom in the PN model. a) Atomic rows
on one side of slip plane (dashed line) are labeled by index \emph{i}.
Distance between adjacent rows in defect-free crystal is $\Delta x$.
b) Displacements of rows with respect to those on other side of slip
plane are described by vector $\vec{\delta}_{i}$.}

\end{figure}

Within the semi-discrete PN framework, a straight dislocation lying
along the $y$-direction is represented in terms of a set of misfit
vectors $\vec{\delta_{i}}$ that describe the disregistry of atomic
row $i$ relative to their counterparts on the other side of the glide
plane, as depicted in Fig.\ref{fig:Fig1}. Panel a) shows a schematic
representation of two sets of atomic rows extending along the $y$-direction,
one set on each side of the glide plane. The rows above the plane
are labeled by the index $i$. Panel b) shows a top view, depicting
a disregistry vector $\vec{\delta_{i}}$ for row $i$. In this manner,
the total misfit associated with a given dislocation, described by
the Burgers vector $\vec{b},$ is thought of as distributed among
the atomic rows along the $x$-axis, subject to the boundary conditions
$\vec{\delta}_{-\infty}=0$ and $\vec{\delta}_{\infty}=\vec{b}$ .
Here, we consider two-dimensional misfit vectors $\vec{\delta_{i}}=\delta_{i}^{e}\hat{x}+\delta_{i}^{s}\hat{y}$,
with edge and screw components along the $x$ and $y$ directions,
respectively. The equilibrium structure of the dislocation is then
represented by that particular misfit distribution $\vec{\delta_{i}}$
that minimizes the dislocation energy per unit length,

\begin{equation}
U_{\mathrm{disl}}=U_{\mathrm{elastic}}+U_{\mathrm{misfit}}+U_{\mathrm{stress}}+C,\label{eq:1}\end{equation}
 where

\begin{eqnarray}
U_{\mathrm{elastic}} & = & \underset{i,j}{\sum}\chi_{ij}\left(K_{e}{\scriptstyle {\textstyle \,\rho}}_{i}^{e}\,\rho_{j}^{e}+K_{s}\,\rho_{i}^{s}\,\rho_{j}^{s}\right),\label{eq:elastic}\\
U_{\mathrm{misfit}} & = & \underset{i}{\sum}\gamma(\vec{\delta}_{i})\Delta x,\label{eq:misfit}\\
U_{\mathrm{stress}} & = & {\scriptstyle \frac{1}{2}}\underset{i}{\sum}\left[\tau^{e}\left(\delta_{i}^{e}+\delta_{i-1}^{e}\right)+\tau^{s}\left(\delta_{i}^{s}+\delta_{i-1}^{s}\right)\right]\Delta x,\label{eq:stress}\end{eqnarray}
and $C$ is a constant that can be ignored \cite{Bulatov1997}. In
the elastic part of the dislocation energy, Eq.~\prettyref{eq:elastic},
$\chi_{ij}$ is a discretized universal kernel \cite{Lu2000}, and
${\scriptstyle {\textstyle {\textstyle K}_{e}}{\textstyle =\mu/4\pi(1-\nu)}}$
and ${\scriptstyle {\textstyle K_{s}=\mu/4\pi}}$ are elastic pre-factors
with $\mu$ the shear modulus and $\nu$ the Poisson's ratio. In addition,
${\textstyle {\scriptstyle {\textstyle \rho_{i}^{e}\equiv}{\textstyle \left(\delta_{i}^{e}-\delta_{i-1}^{e}\right)/\Delta}{\textstyle x}}}$
and ${\scriptstyle {\textstyle \rho_{i}^{s}\equiv\left(\delta_{i}^{s}-\delta_{i-1}^{s}\right)/\Delta x}}$,
where $\Delta x$ is the distance between adjacent rows in the defect-free
crystal. Eq.\prettyref{eq:misfit} represents the misfit contribution,
in which ${\scriptstyle {\textstyle \gamma(\vec{\delta})}}$ is known
as the generalized stacking-fault (GSF) energy surface \cite{Vitek1968}.
It describes the excess energy per unit area of a crystal that is
subjected to the following procedure. It is first cut into two defect-free
parts across a given plane. The two parts are then displaced relative
to each other by a vector $\vec{\delta}$, after which they are patched
together again. An example configuration of the GSF on the basal plane
of the \emph{hcp} structure is the intrinsic stacking fault (ISF),
in which the displacement vector $\vec{\delta}$ describes the associated
shift in the planar stacking. In the context of the PN model, the
GSF surface reflects the inter-atomic interactions in the system and
serves to model the details of the dislocation core on the atomic
scale. Finally, the stress term of Eq.~\prettyref{eq:stress} accounts
for the work done by any external stresses, where $\tau^{e}$ and
$\tau^{s}$ denote the magnitude of the components of the stress tensor
that couple to the edge and screw displacements, respectively \cite{Hirth1982}.
The quantities that specify the model for dislocations in a particular
material are the elastic parameters $\mu$ and $\nu$ , the GSF surface
${\scriptstyle {\textstyle \gamma(\vec{\delta})}}$ associated with
the glide plane of interest, and $\Delta x$.

Here, we employ the SWF model based on the parameter set of Ref.\cite{MacFarland1994}
to determine these quantities for solid \emph{hcp} $^{\text{4}}$He
(space group 194). In order to determine its elastic properties, we
employ a computational cell containing 720 particles at a density
of 0.0294 \AA$^{-3}$, which corresponds to lattice parameters \emph{a}=3.63668
\AA \, and c=5.93866 \AA, subject to standard periodic boundary
conditions. Sampling configurations according to the quantum-mechanical
probability density of the SWF model using the Metropolis algorithm,
we then compute expectation values of the stress tensor \cite{Ceperley1977}
associated with the six independent deformations of the periodic cell,
imposing strain levels of 0.25\%. Using the standard relationship
between the stress and strain tensors \cite{Nye1985}, we extract
the five independent elastic constants of the \emph{hcp} structure.
The results, which, to the best of our knowledge represent the first
complete estimate of the elastic constants in \emph{hcp} $^{\text{4}}$He,
are reported in Table \ref{tab:Table1}. The shear modulus $\mu=C_{44}=17.1\pm0.8$
MPa, is in good agreement with the value of 14 MPa that follows from
the ratio of the experimental shear stress and strain values reported
in \cite{Syshchenko2009}. Poisson's ratio, obtained from the results
in Table \ref{tab:Table1}, is found to be $\nu=0.151.$ Both theoretical
values are those corresponding to shear directions in the basal plane,
in which hexagonal crystals are isotropic \cite{Li1976}.

\begin{figure}[H]
\includegraphics[width=9cm]{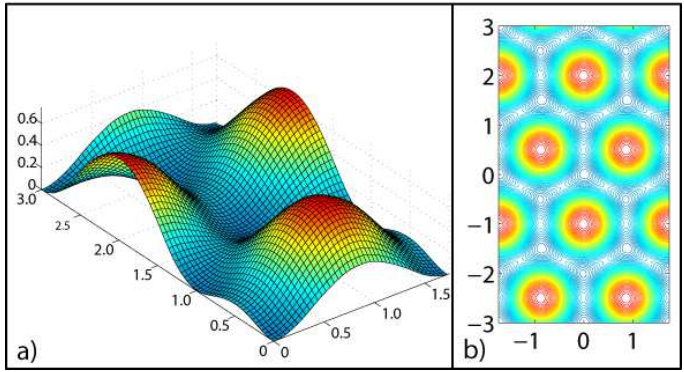}

\caption{\label{fig:Fig2}Fourier series representation of GSF surface on the
basal plane of \emph{hcp} $^{4}\textrm{He}$ as modeled by the SWF
model. a) GSF energy (in mJ/m$^{2}$) as a function of two-dimensional
displacement $\vec{\delta}$ (in units of $b_{p}=\frac{1}{3}\sqrt{3}a=2.0996$
\AA). b) Contour plot of GSF surface reflects symmetry of basal plane
of \emph{hcp} surface\emph{.}}

\end{figure}

Since basal slip is known to be the dominant dislocation glide mechanism
in \emph{hcp} solid $^{4}\mathrm{He}$ \cite{Paalanen1981} we focus
on the properties of these particular dislocations and compute the
GSF surface associated with the basal plane. For this purpose, we
utilize the 720-atom cell and impose a series of 400 slip vectors
$\vec{\delta}$ in the basal plane by adjusting the periodic boundary
condition along the \emph{c}-axis. The shadow degrees of freedom of
the atoms immediately adjacent to the slip plane are allowed to vary
only along the \emph{c}-direction to maintain the relative displacement.
Using the Metropolis algorithm we then sample configurations according
to the SWF and compute the expectation value of the Hamiltonian as
a function of $\vec{\delta}$. Subtracting the expectation value at
$\vec{\delta}=0$ and dividing by the area, we then obtain the GSF
surface. In order to implement the results into the PN mode, we fit
the results using a Fourier series that reflects the lattice symmetry
of the basal plane of the \emph{hcp} structure: $\gamma(\vec{\delta})=\underset{\vec{\mathbf{G}}}{\sum}c_{\vec{\mathbf{G}}}\exp(i\mathbf{\vec{G}}\cdot\vec{\delta})$,
in which we use a set of 81 two-dimensional reciprocal lattice vectors
$\mathbf{\vec{G}}$. The results are shown in Fig.~\ref{fig:Fig2}.
The perfect crystal configuration, which has GSF value of zero, is
associated with the displacement $\vec{\delta}=0$ and its periodic
equivalents. The ISF configuration, which corresponds to the displacement
vector (and equivalents) $\vec{\delta}=(0,\, b_{p})$, with $b_{p}=\frac{1}{3}\sqrt{3}a=2.0996$
\AA \, the length of a partial Burgers vector, has an excess energy
of 0.0063 mJ/$\mathrm{m}^{2}$.

\begin{figure}[H]
\includegraphics[width=9cm]{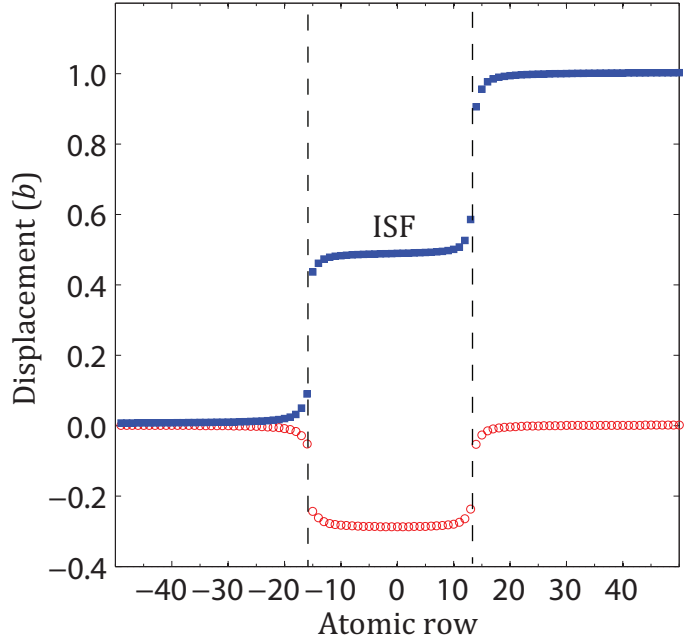}

\caption{\label{fig:Fig3}Optimized displacement profile $\vec{\delta_{i}}$
for screw dislocation on basal plane: edge component (open circles)
and screw component (filled squares) are measured in units of the
Burgers vector $b=a=3.63668$\AA. Dashed lines indicate positions
of partial dislocations. }

\end{figure}

Using our estimates for the elastic properties and the GSF surface
in the PN model, we investigate the structure and intrinsic mobility
of 4 dislocation types on the basal plane: (i) screw, (ii) $30^{\circ}$,
(iii) $60^{\circ}$, and (iv) edge. Fig.~\ref{fig:Fig3} shows the
optimized disregistry profile for the screw dislocation, obtained
by minimizing Eq.~\prettyref{eq:1} at zero external stress. As expected,
given the low stacking-fault energy (SFE) value , it dissociates into
two $30^{\circ}$partial dislocations with opposite edge components,
separated by an ISF area with a width of 29 atomic rows, which corresponds
to 91.33 \AA. The core width $\varsigma$ of the partials, defined
as the distance over which the displacement changes from $\frac{1}{4}$
to $\frac{3}{4}$ of it total value\cite{Lu2000}, is approximately
1 atomic row or $\sim$3.1 \AA. The second line of Table~\ref{tab:Table2}
contains the dissociation widths of the other three dislocations,
showing an increasing ISF width with increasing edge component, consistent
with dislocation theory \cite{Hirth1982}.

In addition to the structural properties described here, the PN model
permits an estimate of the intrinsic dislocation mobility, which is
measured in terms of the Peierls stress. To this end, we impose an
external shear stress in the glide plane parallel to the total Burgers
vector. It produces maximal force per unit length \cite{Hirth1982}
on the dislocation line for the given stress magnitude. This magnitude
is then increased in small steps, followed by minimization of Eq.~\prettyref{eq:1}
with respect to the disregistry vectors $\vec{\delta_{i}}$ . At a
critical stress value, the so-called Peierls stress, an instability
is reached and an equilibrium solution ceases to exist.\cite{Lu2000,Lu2003}
In this situation the dislocation becomes free to move through the
crystal. The third line of Table \ref{tab:Table2} contains the Peierls
stress values for the four considered dislocation types. The lowest
Peierls stress value, obtained for the 30$^{\circ}$ dislocation,
is of the order of 1.5$\times10^{-2}\ \mathrm{MPa}$. This value is
$\sim3$ orders of magnitude smaller than the shear modulus, which
is consistent with the typical discrepancy between the ideal shear
strength and actual yield stresses in crystals \cite{Hirth1982}.

The above results were obtained using the SWF model, which, while
providing a good description of the shear modulus for \emph{hcp} solid
$^{4}\mathrm{He}$, significantly underestimates a recent experimental
estimate for the SFE, $(0.07\pm0.02)\mathrm{\, mJ/}\mathrm{m}^{2}$\cite{Junes2008}.
In order to explore the possible influence of the SFE value on the
dislocation mobility, we repeat the Peierls stress calculations for
the case in which the equilibrium dissociation widths of the four
dislocation types is reduced by a factor 10, which is the ratio between
the experimental and theoretical SFE values. To this end, we apply
an additional shear stress component, whose direction in the glide
plane is perpendicular to the total Burgers vector of the dislocation.
This stress component, known as Escaig stress \cite{Bulatov2006},
does not produce a force on the dislocation as a whole, but mimics
a situation with a different SFE value. Using an Escaig stress of
0.4 MPa, we recompute the Peierls stress values for the four dislocations
types in the basal plane. The results are shown in the fourth row
of Table \ref{tab:Table2}. The effect of an increased effective SFE
value does not significantly affect the Peierls stress values of the
model. This is consistent with earlier PN calculations in metals,
in which the Peierls stress was not found to be sensitive to dissociation
width \cite{Lu2000}.

Finally, we examine our results in the context of recent experiments
considering the macroscopic behavior of solid $^{\text{4}}\mathrm{He}$
at low temperatures. In the observation\cite{Kim2004a,Kim2004} of
non-classical rotational inertia (NCRI), interpreted as a signature
of superfluidity, the crucial role of crystal defects and disorder
seems firmly established. Specifically, the behavior of dislocations
has attracted a particular interest after the discovery of an unexpected
increase of the shear modulus that shows the same temperature and
$^{\text{3}}$He impurity concentration dependence as the original
NCRI observations \cite{Day2007,Day2009,West2009a}. Inspired by the
continuum-elasticity based Granato-L{\"u}cke theory \cite{Granato1956},
it has been hypothesized that this stiffening is a consequence of
a change of mobility of a network of dislocations. This network is
thought to be pinned by $^{\text{3}}$He impurities at lowest temperatures,
while it becomes mobile under warmer conditions. Analyzing the dislocation
mobility results of our model, it is interesting to observe that our
lowest Peierls barrier is about 20 times \emph{larger} than the shear
stresses of $\sim700$ Pa reached in recent experiments \cite{Day2009}.
This suggests that intrinsic lattice resistance is an essential factor
when it comes to the mobility of dislocations on the basal plane in
\emph{hcp} solid $^{4}\mathrm{He}$. Indeed, at the stress levels
reported in these recent experiments, such dislocations would not
be expected to be mobile, not even in the absence of any pinning centers.
Moreover, it is not expected that the Peierls stress varies significantly
as a function of temperature below 0.1 K, at which the stiffening
is observed, given that finite temperature path-integral Monte Carlo
calculations of several properties do not show a significant temperature
dependence below 1 K \cite{Ceperley1995}. In this context, a satisfactory
explanation for the observed elastic stiffening in \emph{hcp} solid
$^{4}\mathrm{He}$ must involve intrinsic mobility issues.

In summary, we employ a hybrid continuum-atomistic approach, based
on the Peierls-Nabarro model and the shadow wave function formalism,
to obtain direct insight into the intrinsic structural and mobility
properties of dislocations in \emph{hcp} solid $^{4}\mathrm{He}$
at zero temperature. In addition to providing key information concerning
the elastic properties of this prototypical quantum crystal, the results
reveal a significant lattice resistance to dislocation motion. Analyzing
our results in the context of the proposed dislocation-pinning interpretation
of the similarity between the NCRI and elastic stiffening phenomena
suggests that intrinsic lattice resistance is an essential factor
when it comes to the mobility of dislocations. The proposed interpretation,
which entirely ignores this element, may therefore not provide a satisfactory
explanation for the observed elastic stiffening in \emph{hcp} solid
$^{4}\mathrm{He}$.

The authors acknowledge financial support from the Brazilian agencies
FAPESP, CNPq and CAPES. Part of the computations were performed at
the CENAPAD high-performance computing facility at Universidade Estadual
de Campinas.

\bibliographystyle{apsrev}

\begin{table}[H]
\caption{\label{tab:Table1}Elastic constants of the SWF mode\emph{l for hcp}
$^{\text{4}}$He within the SWF model, in units of MPa.}

\begin{tabular}{|c|c|c|c|c|}
\hline 
$C_{11}$  & C$_{33}$  & C$_{44}$  & C$_{12}$  & C$_{13}$\tabularnewline
\hline
\hline 
60.8$\pm0.8$  & 77.9$\pm0.8$  & 17.1$\pm0.8$  & 34.4$\pm0.8$  & 14.4$\pm0.8$\tabularnewline
\hline
\end{tabular}
\end{table}

\begin{table}[H]
\caption{\label{tab:Table2}Core structure and Peierls Stress for four dislocations
on the basal plane for the SWF model for solid \emph{hcp} $^{4}\mathrm{He}$.}

\begin{tabular}{|c|c|c|c|c|}
\hline 
 & Screw  & $30^{\circ}$  & $60^{\circ}$  & Edge\tabularnewline
\hline
\hline 
Partial separation (\AA)  & 91.3  & 109.1  & 119.7  & 127.3\tabularnewline
\hline 
Peierls Stress (MPa)  & 0.41$\pm$0.01  & 0.015$\pm$0.01  & 0.23$\pm$0.01  & 0.08$\pm0.01$\tabularnewline
\hline 
Peierls Stress corrected SFE (MPa)  & 0.48$\pm0.01$  & 0.045$\pm0.01$  & 0.39$\pm0.01$  & 0.08$\pm0.01$\tabularnewline
\hline
\end{tabular}
\end{table}

\end{document}